\documentclass[%
 preprint,
 superscriptaddress,
 amsmath,amssymb,
 aps,
 showkeys,
 titlepage,
 endfloats*   
]{revtex4-2}

\newcommand{\code}[1]{\texttt{#1}}
\usepackage[utf8]{inputenc}
\usepackage[english]{babel}
\usepackage{chemformula}
\usepackage[version=4]{mhchem}
\usepackage{siunitx}
\usepackage{graphicx}
\usepackage{dcolumn}
\usepackage{bm}
\usepackage[mathlines]{lineno}
\usepackage[colorlinks = true,
            linkcolor = blue,
            urlcolor  = blue,
            citecolor = red,
            anchorcolor = blue]{hyperref}

\begin{document}

\title{Extraordinary Thermoelectric Properties of Topological Surface States in Quantum-Confined \ce{Cd3As2} Thin Films} 

\author{Wenkai Ouyang}
\affiliation{Department of Mechanical Engineering, University of California, Santa Barbara, CA 93106, USA}

\author{Alexander C. Lygo}
\affiliation{Materials Department, University of California, Santa Barbara, CA 93106, USA}

\author{Yubi Chen}
\affiliation{Department of Mechanical Engineering, University of California, Santa Barbara, CA 93106, USA}
\affiliation{Department of Physics, University of California, Santa Barbara, CA 93106, USA}

\author{Huiyuan Zheng}
\affiliation{Department of Physics, the University of Hong Kong, Hong Kong, 999077, China}

\author{Dung Vu}
\affiliation{Department of Mechanical and Aerospace Engineering, the Ohio State University, Columbus, OH 43210, USA}

\author{Brandi L. Wooten}
\affiliation{Department of Mechanical and Aerospace Engineering, the Ohio State University, Columbus, OH 43210, USA}

\author{Xichen Liang}
\affiliation{Department of Chemical Engineering, University of California, Santa Barbara, CA 93106, USA}

\author{Wang Yao}
\affiliation{Department of Physics, the University of Hong Kong, Hong Kong, 999077, China}

\author{Joseph P. Heremans}
\email{heremans.1@osu.edu}
\affiliation{Department of Mechanical and Aerospace Engineering, the Ohio State University, Columbus, OH 43210, USA}

\author{Susanne Stemmer}
\email{stemmer@ucsb.edu}
\affiliation{Materials Department, University of California, Santa Barbara, CA 93106, USA}

\author{Bolin Liao}
\email{bliao@ucsb.edu} \affiliation{Department of Mechanical Engineering, University of California, Santa Barbara, CA 93106, USA}

\date{\today}

\begin{abstract}
Topological insulators and semimetals have been shown to possess intriguing thermoelectric properties promising for energy harvesting and cooling applications. However, thermoelectric transport associated with the Fermi arc topological surface states on topological Dirac semimetals remains less explored. In this work, we systematically examine thermoelectric transport in a series of topological Dirac semimetal \ce{Cd3As2} thin films grown by molecular beam epitaxy. Surprisingly, we find significantly enhanced Seebeck effect and anomalous Nernst effect at cryogenic temperatures when the \ce{Cd3As2} layer is thin. Combining angle-dependent quantum oscillation analysis, magnetothermoelectric measurement, transport modelling and first-principles simulation, we isolate the contributions from bulk and surface conducting channels and attribute the unusual thermoeletric properties to the topological surface states. Our analysis showcases the rich thermoelectric transport physics in quantum-confined topological Dirac semimetal thin films and suggests new routes to achieving high thermoelectric performance at cryogenic temperatures.  

\end{abstract}

                            
\maketitle



\section{Introduction}

Topological insulators and semimetals have recently attracted intense research interests due to their exotic bulk transport properties, such as the chiral anomaly and topologically protected surface states~\cite{hasan2010colloquium,lv2021experimental}.
Additionally, their unusual thermal and thermoelectric transport properties have also offered opportunities for potential thermoelectric energy conversion applications~\cite{fu2020topological,vu_ThermalChiralAnomaly}. 
For example, topological insulators and topological semimetals typically feature linear or close-to-linear electronic bands near the Fermi level, leading to a high charge carrier mobility~\cite{schumann_MBE} that is beneficial for thermoelectrics. 
It has been experimentally shown that the thermoelectric performance of lead selenide (PbSe) can be enhanced by a pressure-driven topological phase transition, where the formation of linear Dirac bands improves the electrical conductivity~\cite{chen2019enhancement}. Furthermore, both theory~\cite{skinner2018large} and experiments~\cite{han2020quantized,zhang2020observation} have established that the Seebeck coefficient of topological semimetals can increase with the magnetic field in the quantized regime without saturation. 
On the thermal transport side, first-principles simulations and inelastic neutron and x-ray scattering measurements have suggested significant phonon softening in topological semimetals due to the Kohn anomaly, leading to reduced thermal conductivity~\cite{yue_SoftPhonons,yue2020phonon,nguyen2020topological}. 
These properties are all promising features for improving thermoelectric energy conversion efficiency. 
So far, however, the explored thermoelectric properties of topological insulators and semimetals are mostly limited to those in intrinsic bulk crystals, while epitaxial thin films remain less examined. 
Compared with bulk crystals, epitaxial thin films offer more degrees of control, such as strain and quantum confinement, which can be further tuned to optimize their thermoelectric properties. In addition, topological surface states also contribute more prominently in thin films. 
In this work, we focus on detailed measurements and modeling of thermoelectric properties of epitaxial thin films of a topological Dirac semimetal \ce{Cd3As2}, aiming to isolate the impacts of surface states and quantum confinement.    

\ce{Cd3As2} is an archetypal topological Dirac semimetal that exhibits a three-dimensional linear dispersion relation and topologically protected surface states \cite{crassee_review, wang_3DCd3As2,yangClassificationTopo,chorsi2020widely}. 
The linear dispersion relation and a large Fermi velocity give rise to an electron mobility as high as 20,000 $\mathrm{cm^2}$/V-s at room temperature~\cite{schumann_MBE} in thin films grown by molecular beam epitaxy (MBE). 
In addition, \ce{Cd3As2} has a lattice thermal conductivity ~\cite{spitzerTC,armitageTC,wang_thermoelectric} around 0.3 $\sim$ 0.7\,W/m-K at 300\,K, which is anomalously low for single crystals. 
Using first-principles lattice dynamics calculations, Yue \textit{et al.} \cite{yue_SoftPhonons} uncovered the presence of a cluster of soft optical phonon modes in \ce{Cd3As2}, which significantly enhance the scattering phase space of acoustic phonons leading to a low lattice thermal conductivity. 
The combination of a high electron mobility and a low thermal conductivity makes it a highly promising candidate for thermoelectric applications. 
However, as a topological Dirac semimetal, \ce{Cd3As2} has no bandgap in its bulk form, which is typically detrimental to the Seebeck coefficient. As a result, a moderate Seebeck coefficient ranging from 50 $\mu \mathrm{V/K}$ to 150 $\mu \mathrm{V/K}$ \cite{pariari_TuningScatteringMechanism,jia_ThermoelectricChiral,wang_thermoelectric,amarnath_theory,liang_AnomalousNernstEffect} was recorded experimentally in \ce{Cd3As2} bulk crystals.

To overcome this limitation, a bandgap can be opened in \ce{Cd3As2} through either crystal symmetry breaking or quantum confinement that can be realized in epitaxially grown thin films or heterostructures~\cite{wang_3DCd3As2, youngDirac3D, panTopologicalControl,xiaoQuantumConfinement,narayanTopologicalTuning}. Specifically, it has been shown experimentally that in \ce{Cd3As2} thin films grown by MBE, a bulk band gap emerges below a certain film thickness due to quantum confinement, where the low temperature electrical transport properties are dominated by mobile carriers restricted to two-dimensional surface states~\cite{schumann_observation}. The surface states in topological Dirac semimetal \ce{Cd3As2} consists of two sets of Fermi arcs~\cite{arribi2020topological} [except for the (001) surface where the projection of the bulk Dirac nodes becomes a point] that are capable of producing quantum oscillations and the quantum Hall effect ~\cite{potterQuantumOscillationsSurface,wang3DQuantumHall}, which has been experimentally observed in MBE-grown films with a thickness below 70\,nm~\cite{schumann_observation, goyal_ThicknessDependence}. These surface states have been theoretically predicted to persist even in thin films with a few nm thickness~\cite{arribi2020topological}. Furthermore, the Dirac dispersion of surface states on (112) surfaces in MBE-grown \ce{Cd3As2} has been verified through gating experiments~\cite{galletti2018two}. However, thermoelectric transport properties of these topological surface states are less explored and the impact of quantum confinement on thermoelectric properties has not been examined experimentally.

In this work, we systematically investigate the thermoelectric transport in high-quality \ce{Cd3As2} thin films grown by MBE. To resolve the contribution from different conducting channels, we studied samples with varying \ce{Cd3As2} thicknesses: 950\,nm in the bulk limit, 95\,nm in the transition regime, and 25\,nm with quantum confinement and prominent contributions from the surface states. Surprisingly, we observed anomalously enhanced Seebeck and Nernst coefficients in the 25\,nm sample at cryogenic temperatures. Combining extensive magneto-Seebeck and Nernst effect measurements, detailed angle-dependent quantum oscillation characterization, analytical transport models and first-principles density functional theory (DFT) simulations, we attribute the observed thermoelectric properties to the interplay of bulk and surface conducting channels in parallel. Our work provides a comprehensive understanding of thermoelectric transport in confined topological Dirac semimetal thin films and reveals the rich physics in these systems that can be potentially employed to engineer composite materials for thermoelectric applications.

\section{Results and Discussion}

\subsection{Temperature dependence of resistivity and Seebeck coefficient}

The MBE technique was employed to achieve the epitaxial growth of (112)-oriented \ce{Cd3As2} thin films on (111)-oriented GaAs substrates with a 150-nm (111) \ce{GaSb} buffer layer~\cite{schumann_MBE}. More details of the growth method and the measurement technique can be found in the Methods section. Characterization results of the bare \ce{GaSb} buffer layer is provided in Supplementary Note 1 in the Supplementary Information (SI). Fig.~\ref{fig:fig1}a presents a schematic illustration of the sample structures studied in this work as well as a picture of the measurement device. 
In our experiment, three different thicknesses of \ce{Cd3As2} were prepared: 950\,nm, 95\,nm and 25\,nm, with the same GaSb buffer layer and substrate conditions. We refer to them as ``950 nm sample'', ``95 nm sample'', and ``25 nm sample'', respectively, in the following discussion.

The measured electrical resistivity of the three samples as a function of temperature is shown in Fig.~\ref{fig:fig1}b.  The resistivity of the 950-nm sample exhibits a typical semimetallic behavior, where it increases with temperature. This behavior reflects the bulk \ce{Cd3As2} property. In contrast, the 95-nm sample shows a more pronounced semiconductor-like behavior with a decreasing resistivity with temperature. Low-field Hall results of the two samples are provided in the Supplementary Note 2 in the SI, indicating n-type conduction that originates from the bulk \ce{Cd3As2} states. A very high Hall mobility up to Therefore, the semiconductor-like behavior in the 95-nm sample suggests the onset of bulk bandgap opening due to either weak quantum confinement or strain that breaks the four-fold rotation symmetry around (001) in \ce{Cd3As2} protecting the Dirac nodes~\cite{arribi2020topological}, where the thermal activation of carriers results in a decrease in the resistivity with an increasing temperature~\cite{galletti_NitrogenPassivation, jia_ThermoelectricChiral}. For the 25-nm sample, a modest variation in resistivity is noted with an increasing temperature, suggesting a delicate interplay between the semiconductor-like behavior of the bulk states and, potentially, the metallic surface states \cite{galletti_NitrogenPassivation}.


In Fig.~\ref{fig:fig1}c, we present the temperature-dependent Seebeck coefficient for the three samples. Specifically, for the 950-nm sample, we observed an increase in the Seebeck coefficient with temperature, where the negative sign suggests the prevalence of n-type carriers \cite{liang_AnomalousNernstEffect, jia_ThermoelectricChiral}. Meanwhile, for the 95-nm sample, the Seebeck coefficient exhibits a similar temperature dependence above 50\,K. In the temperature range above 50\,K, the Seebeck coefficient of both samples can be well described by our DFT simulation using the bulk band structure of \ce{Cd3As2} assuming a Fermi level slightly into the conduction band (details of the DFT simulation are given in Methods and Supplementary Note 3 in the SI). However, around 50\,K, a change in the sign of the Seebeck coefficient occurs, suggesting a transition of the dominant carriers from n-type to p-type. At even lower temperatures, we also observed an increase of the positive Seebeck coefficient. Both features cannot be explained by our DFT simulation considering only the \ce{Cd3As2} bulk states. More interestingly, for the 25-nm sample, a positive Seebeck coefficient was observed in the entire temperature range, and the Seebeck coefficient showed a metallic feature with a negative temperature dependence above 50\,K, contrary to the behavior observed in the other two samples. Strikingly, the Seebeck coefficient displayed a sharp increase below 50\,K, peaked at 5\,K with a value of nearly 500 $\mu \mathrm{V/K}$. This surprising behavior was in sharp contrast to the other two samples with thicker \ce{Cd3As2} thin films. Furthermore, the distinct behaviors of the Seebeck coefficient in the three samples suggest the existence of contributions from more than one conducting channels in the thinner samples, thereby requiring further investigations as detailed in subsequent sections.



\subsection{Angle-dependent quantum oscillations} \label{subsec:a}
In order to comprehensively investigate the underlying mechanisms responsible for the distinct thermoelectric behaviors of the three heterostructure samples with different \ce{Cd3As2} thicknesses, we performed systematic measurements of angle-dependent quantum oscillations in both the electrical resistivity (the Shubnikov-de-Haas, or SdH, oscillation) and the Seebeck coefficient of the three samples at 2\,K. Quantum oscillations of electrical transport properties as a function of the magnetic field are manifestations of the Fermi surface quantization, where the oscillation frequency as a function of the angle of the magnetic field can provide detailed information of the Fermi surface ~\cite{liang2015ultrahigh,he_QuantumOscillation, ashcroft1976solid}. Figure ~\ref{fig:fig2}a presents the Seebeck voltage oscillation of the three samples when the magnetic field is perpendicular to the sample surface (``0 degree'' direction) at 2\,K. The Seebeck voltage is plotted against the inverse magnetic field ($1/B$), from which the oscillation frequencies are obtained through a Fourier transform (FT). The three samples exhibit distinct oscillation frequencies, signaling different origins of the observed thermoelectric transport~\cite{xiangAngularDependentCd3As2, he_QuantumOscillation, goyal_SurfaceStatesStrained}. Specifically, for the 950 nm sample, a single prominent oscillation frequency is detected at approximately 20\,T, as indicated by peaks A-B labeled in Fig.~\ref{fig:fig2}a. The 95 nm sample shows two oscillation frequencies: one is around 20\,T, also denoted by peaks A-B, while the other is around 55\,T, corresponding to peaks C-D in Fig.~\ref{fig:fig2}a. Similarly, the 25-nm sample also displays these two oscillation frequencies.  

To further pinpoint the conducting channels corresponding to these quantum oscillation frequencies, we conducted a systematic investigation of the SdH and Seebeck coefficient quantum oscillation frequencies as a function of the angle of the applied magnetic field in all three samples. Figures~\ref{fig:fig2}b and~\ref{fig:fig2}d show the results and a comparison of these results with the DFT calculation considering only the bulk \ce{Cd3As2} states. According to the Onsager relation~\cite{etoSdH}, the oscillation frequencies are proportional to extremal areas of the Fermi surface perpendicular to the applied magnetic field. Figure~\ref{fig:fig2}b illustrates the variation of the SdH oscillation frequency as a function of the magnetic field angle in 950-nm and 95-nm samples, respectively. The oscillation frequency is observed to undergo slight variations with angle in both samples, with the 950 nm sample exhibiting a concave shape centered around 90$^{\circ}$ and the 95 nm sample a convex shape. From Fig.~\ref{fig:fig2}b, our DFT calculation predicts an oscillation frequency around 16\,T at zero degree field angle, assuming bulk \ce{Cd3As2} states and a Fermi level $\sim 12$\,meV above the charge neutral point (CNP). The Fermi level position is deduced from the Hall carrier concentration measurement of the 950 nm sample (see Supplementary Note 3 in the SI). Additionally, the DFT simulation also reproduces the concave trend shown in Fig.~\ref{fig:fig2}b for the 950 nm sample. Specifically, at the predicted Fermi level position, a Lifshitz transition occurs and the two spherical Fermi surfaces surrounding the two Dirac points along the $\Gamma$-Z direction merge to form a peanut-shaped Fermi surface, as depicted in Fig.~\ref{fig:fig2}c. This Fermi surface shape accounts for the concave trend observed in the 950 nm sample. The agreement between the 950 nm sample data and the DFT simulation further suggests that the thermoelectric transport in the 950 nm sample mainly arises from the n-type \ce{Cd3As2} bulk states. Similarly, the 95 nm sample shows a SdH frequency at the 0$^{\circ}$ field angle in agreement with the DFT calculation, indicating the contribution from the \ce{Cd3As2} bulk states. However, the angle-dependent trend observed in the 95 nm sample deviates from that in the 950 nm sample, which can be potentially attributed to the onset of the quantum confinement effect that can vary the shape of the Fermi surface as well as the position of the Fermi level. Unfortunately, in the 95 nm sample, the additional high-frequency SdH oscillation at 55\,T is weak and we could not map its angle-dependence reliably.  

For the 25 nm sample, it turned out difficult to obtain reliable angle-dependent SdH data with multiple oscillation frequencies. Instead, we found that the quantum oscillation observed in the Seebeck coefficient as a function of the magnetic field is much clearer and provides more reliable extraction of the oscillation frequencies. Therefore, we conducted systematic angle-dependent Seebeck oscillation measurements on the 25 nm sample and the results are shown in Fig.~\ref{fig:fig2}d. One of the observed oscillation frequencies near 20\,T is nearly independent of the magnetic field direction, signaling its origin in the \ce{Cd3As2} bulk states, which are expected to form closely spaced subbands due to quantum confinement. In addition, the 25 nm sample exhibits another high oscillation frequency around 55\,T at 0$^{\circ}$ field direction, which also appears in the 95 nm sample. The angle-dependent reduction in this oscillation frequency culminating at 90$^{\circ}$ field direction implies that this frequency is associated with two-dimensional (2D) surface states, since the Fermi surface associated with the 2D surface states has zero intersection with the magnetic field at 90$^{\circ}$ field direction, as illustrated in Fig.~\ref{fig:fig2}e. The observation of this higher-frequency oscillation and its field-angle dependence is consistent with previous quantum oscillation measurements of \ce{Cd3As2} thin films~\cite{nishihaya2019quantized}. Based on this observation, the behavior of the Seebeck coefficient of the 95 nm sample shown in Fig.~\ref{fig:fig1}c can be understood. Above 50\,K, the main contribution to the Seebeck coefficient in the 95 nm sample comes from the bulk \ce{Cd3As2} states, similarly as the 950 nm sample. Below 50\,K, the bulk states in the 95 nm sample start to freeze out due to the onset of gap opening, while the contribution from the surface states emerges and is responsible for the sign change of the Seebeck coefficient. This view is also consistent with the p-type character of the surface states observed in previous measurements of \ce{Cd3As2} thin films when the surface Fermi level was low, due to either surface depletion or electrostatic gating ~\cite{galletti_NitrogenPassivation,galletti2018two}. In contrast, the Seebeck coefficient of the 25 nm sample is positive throughout the entire temperature range, suggesting that the bulk \ce{Cd3As2} states, presumably n-type, play a minor role in thermoelectric transport even up to 300\,K. Instead, the contribution of the p-type surface states is largely amplified in the 25 nm sample compared with the 95 nm sample due to its reduced thickness and stronger quantum confinement that gaps out the bulk states. In the following, we provide detailed magnetothermoelectric transport measurements and model analysis to further quantify the contribution from surface states.

\subsection{Magnetothermoelectric transport measurements and modeling}
Based on the quantum oscillation analysis, we have established that the thermoelectric transport in the 950-nm sample is dominated by the n-type bulk \ce{Cd3As2} states, while additional contribution from the surface states play a role below 50\,K in the 95-nm sample. In the 25-nm sample, the contribution from the surface states is significant even up to 300\,K. To further verify and quantify our interpretation, we discuss detailed magnetotransport measurements, including Hall, Seebeck and Nernst coefficients, and the associated transport models in this section. We focus on analyzing the 25-nm sample here, while data of the 950-nm sample and the 95-nm sample are provided in the Supplementary Note 2 in the SI. All measurements discussed in this section were done with a magnetic field perpendicular to the sample surface. Firstly, the magnetoresistance and the Hall resistivity data (Supplementary Note 2 in the SI) of the 25 nm sample exhibit a typical two-carrier behavior with contributions from two groups of carriers with opposite charge. In particular, a group of higher-mobility n-type carriers dominate the small-field Hall resistivity, while a group of higher-density p-type carriers dominate the large-field Hall resistivity. To simultaneously incorporate the magnetoresistance and the Hall resistivity, we used a two-band Drude model to fit the Hall resistivity $\rho_{xy}$ data as a function of temperature and magnetic field, as shown in Fig.~\ref{fig:fig3}a, where we presented results at three representative temperatures. Details of the model and additional data are provided in Supplementary Note 2 of the SI. The semiclassical Drude model is valid for low magnetic fields before quantum oscillations appear, so we focused on analyzing the magnetotransport data below 4\,T. We used the carrier concentration and the carrier mobility of both groups of carriers as the fitting parameters here. To increase the fitting sensitivity, we used the same group of parameters to simultaneously fit the low-field ($<$ 1\,T) data of the longitudinal resistivity. As shown in Fig.~\ref{fig:fig3}a, the two-band Drude model fits the experimental data satisfactorily. The fitted mobilities and concentrations for the electron and hole conduction channels are shown in Fig.~\ref{fig:fig3}b and c. As expected, the Hall data can be explained by a high-mobility, low-density electron channel representing the bulk \ce{Cd3As2} states and a low-mobility, high-density hole channel that can be attributed to the surface states~\cite{galletti_NitrogenPassivation}. The low concentration of the n-type \ce{Cd3As2} bulk states suggests a strong quantum confinement. The high density of the surface state carriers is in qualitative agreement with the observed high-frequency quantum oscillation as well as previous reports on similar samples~\cite{galletti_NitrogenPassivation}. 

Figure~\ref{fig:fig4} shows the magnetic-field-dependent Seebeck coefficient $S_{xx}$ and Nernst coefficient $S_{xy}$ in the 25-nm sample measured at different temperatures in a magnetic field up to 9\,T. In the semiclassical regime, the Seebeck coefficient and the Nernst coefficient can be calculated as~\cite{liang_EvidenceMassiveBulk}:
\begin{equation}
S_{xx}(B) = A(\frac{\sigma_{xx}^2}{\sigma_{xx}^2+\sigma_{xy}^2}D+\frac{\sigma_{xy}^2}{\sigma_{xx}^2+\sigma_{xy}^2} D_H),
\label{eqn:seebeck}
\end{equation}
\begin{equation}
S_{xy}(B) = A \frac{\sigma_{xx}\sigma_{xy}}{\sigma_{xx}^2+\sigma_{xy}^2}(D_H-D),
\label{eqn:nernst}
\end{equation}
where $A=\frac{\pi^2 k_B^2 T}{3e}$, and $\sigma_{xx}$ and $\sigma_{xy}$ are longitudinal and Hall conductivity, respectively. Additional parameters $D$ and $D_H$ are from the Mott relations: $D=\frac{\partial \ln \sigma_{xx}}{\partial \zeta}$, $D_H=\frac{\partial \ln \sigma_{xy}}{\partial \zeta}$, where $\zeta$ is the Fermi level. As shown in Fig.~\ref{fig:fig4}a, the measured Seebeck coefficient follows the semiclassical increase and saturation trend with the magnetic field in a semimetal~\cite{liang_AnomalousNernstEffect}. In Fig.~\ref{fig:fig4}b, we used the two-band semiclassical model (Eqn.~\ref{eqn:seebeck}) with parameters extracted from the Hall resistivity fitting (Fig.~\ref{fig:fig3}) to fit the low-field ($<$ 1\,T) Seebeck data~\cite{liang_EvidenceMassiveBulk}. For consistency, we also used the experimental $\sigma_{xx}$ and $\sigma_{xy}$ data and the fitting parameters $D$ and $D_H$ to fit the low-field Seebeck data, where similar values of $D$ and $D_H$ are obtained (see Fig.~S6 in the SI). The fitted values of $D$ and $D_H$ are provided in Fig.~\ref{fig:fig4}c. The extracted $D$ and $D_H$ values steeply rise as the temperature decreases, reaching $>1000$\,eV$^{-1}$, suggesting unusually sharp changes of $\sigma_{xx}$ as a function of the Fermi level. At cryogenic temperatures where the electronic scattering rates are largely independent of the Fermi level, the sharp changes of $\sigma_{xx}$ most likely stem from the large effective mass and, thus, large density of states of the p-type carriers residing in the surface states. This interpretation is consistent with the Hall resistivity data and in agreement with previous Landau level spectroscopy analysis of similar samples~\cite{galletti2018two}.

The unusual thermoelectric properties of the topological surface states are further manifested in the Nernst coefficient of the 25 nm sample, as shown in Figs.~\ref{fig:fig4}d-f, where we provide the measured Nernst coefficient up to 9\,T and the semiclassical (Eqn.~\ref{eqn:nernst}) prediction for the low-field Nernst data with parameters extracted from fitting the low-field Seebeck data. The low-field Nernst coefficient above 40\,K can be reasonably explained using the semiclassical model (Fig~\ref{fig:fig4}e). Below 40\,K, as shown in Fig.~\ref{fig:fig4}f, however, there is significant discrepancy between the experimental values and the model predictions, suggesting the failure of the semiclassical model. For comparison, the Nernst coefficient measured in our 950 nm sample at cryogenic temperatures (Supplementary Note 2 in the SI) is in good agreement with previous measurements of bulk \ce{Cd3As2} crystals~\cite{liang_AnomalousNernstEffect}, where the anomalous Nernst effect (ANE) contribution due to the Berry curvature of bulk \ce{Cd3As2} states is considered. The Nernst coefficient in the 95 nm sample is slightly increased from that in the 950 nm sample, while the Nernst coefficient in the 25 nm sample is about one order of magnitude higher than that in the 950 nm sample. This experimental trend signals a significantly enhanced ANE associated with the topological surface states in the 25 nm sample. Due to the still scarce understanding of the Fermi arc surface states in \ce{Cd3As2} thin films~\cite{arribi2020topological}, it is difficult to quantitatively verify the large ANE associated with them from a modeling point of view. One potential explanation is that the surface states are gapped in the 25 nm sample due to hybridization of the top and bottom surfaces and/or strain~\cite{arribi2020topological}, which can lead to a large Berry curvature~\cite{chen2022anomalous,hosur2013recent} that contributes to an enhanced ANE. Our result exemplifies the complexity of thermoelectric transport in the quantum-confined thin films of topological semimetals and suggests a new route to achieving large thermoelectric responses at cryogenic temperatures via topological surface states.

In passing, we briefly discuss the possible contribution to the measured thermoelectric response from the GaSb buffer layer, which is a semiconductor and can potentially possess a high Seebeck coefficient. Measurements of a bare GaSb buffer layer are provided in the SI, where it is shown that GaSb becomes insulating below 50\,K. In principle, however, the band alignment between GaSb and \ce{Cd3As2} at the interface can lead to charge transfer and ``doping'' of GaSb, increasing its conductivity. However, in extensive previous studies of this system down to very thin \ce{Cd3As2} layers, there is no experimental evidence that the GaSb buffer layer affects transport measurements in any way, especially at cryogenic temperatures~\cite{schumann_observation,galletti2018two,galletti_NitrogenPassivation,lygo2023two}. Moreover, the charge carriers in GaSb do not carry any nontrivial Berry curvature and, thus, are not expected to lead to any contribution to the observed ANE. On the basis of these observations, we attribute the observed unusual thermoelectric properties to the topological surface states, rather than the GaSb buffer layer. 

\section{Conclusion}
In summary, we systematically studied the thermoelectric transport properties of epitaxially grown \ce{Cd3As2} thin films with different thicknesses. Through detailed quantum oscillation analysis and magnetotransport measurement complemented by DFT simulation and a two-band Drude transport model, we identified contributions from both bulk states and topological surface states. In particular, we revealed unusually large Seebeck and Nernst effect associated with the topological surface states. Our results showcase the complexity of thermoelectric transport in topological semimetal thin films and signal new opportunities to achieve remarkable thermoelectric performances in these emerging material systems.

\section{Methods}
\subsection{Molecular Beam Epitaxy Growth}
(112)-oriented \ce{Cd3As2} films with thicknesses of 25\,nm, 95\,nm, and 950\,nm were grown on 150-nm GaSb buffer layers on GaAs (111)B substrates by molecular beam epitaxy, with similar parameters as described previously~\cite{schumann_MBE}. The GaSb buffer layer was grown with a III/V beam equivalent pressure (BEP) ratio of 4.5 at a substrate temperature of 420\textdegree C. Following the GaSb growth, the substrate temperature is reduced to 135\textdegree C for the Cd$_3$As$_2$ growth. Cd$_3$As$_2$ flux is supplied from molecular sources at a BEP between $2 \times 10^{-7}$ and $5 \times 10^{-6}$ Torr.

\subsection{Thermoelectric Transport Measurement}
In this study, we utilized a steady-state transport measurement system to measure the electrical and thermoelectric properties. Specifically, a commercial 9\,T Dynacool Physical Property Measurement System (PPMS, Quantum Design) was used. For the resistivity and the Hall measurements, we conducted six-point probe measurements with the magnetic field perpendicular to the sample (device pictures are shown in the SI). Voltage measurements were carried out using Keithley 2182A nanovoltmeters, while the power supply was provided by a Keithley 6221 current source. The formation of electrical contact was facilitated using silver epoxy. For thermoelectric measurements, temperature readings were obtained using 25\,$\mu$m T-type thermocouples fabricated through spot welding to minimize parasitic heat loss. It should be noted that the thermocouples' sensitivity deminishes below 60\,K, where Cernox thermistors were employed instead. To detect the Seebeck and Nernst voltage, longitudinal and transverse copper voltage probes were used. A 120-$\Omega$ strain gauge served as the heater, and the heating power was adjusted accordingly to achieve a temperature rise of 1~$\sim$~3 $\%$. \ce{Cd3As2} samples were placed on sapphire heat sinks. The magnetic field was swept from -9\,T to 9\,T with a sweeping rate of 5\,mT/s and a customized LabVIEW control code was used to effectively manage experimental control and the data acquisition process.

\subsection{Angle-dependent Quantum Oscillation Measurement}
A picture of the experimental setup using a rotation sample stage is provided in the SI. At a rotation angle of 0 degree, the heat flow or current flow was applied along the sample, while the magnetic field was perpendicular to the sample. Angle-dependent measurements taken at 15-degree intervals between 0 and 180 degrees. The magnetic field was swept from -9\,T to 9\,T, and measurements included SdH oscillations, as well as thermoelectric oscillations (i.e. oscillations in Seebeck and Nernst coefficients at higher magnetic fields). The raw data was fitted using polynomial equations to establish a baseline, which were subsequently subtracted to isolate the oscillations. Following the subtraction, Fast Fourier Transform (FFT) analysis were performed to extract the oscillation frequencies. The FFT range selected for analysis was from 2\,T to 8.5\,T, as lower field values lead to no discernible oscillations and high noise was observed near 9\.T. For the 25-nm \ce{Cd3As2} sample, a comparative analysis was conducted on both Seebeck and Nernst oscillation signals to ensure the presence of the same oscillation frequencies in both measurements.

\subsection{Density Functional Theory Simulation}
We employed the Vienna ab initio simulation package (\code{VASP})~\cite{kresse1996vasp1, kresse1996vasp2} version 6.3.1. for DFT-based first principles calculations. 
The projector augmented wave (PAW)~\cite{blochl1994projector} potentials were utilized. 
The valence electrons for Cd included $5s^23d^{10}$, while for As they were $4s^23p^3$ electrons. 
The Perdew-Burke-Ernzerhof (PBE)~\cite{perdew1996generalized} exchange correlation functional was adopted, including the spin-orbit coupling.
 The cutoff energy for the plane-wave expansion was 400\,eV. 
The lattice structure was optimized with a Hellmann-Feynman force tolerance of 0.001\,eV/Å. 
The Monkhorst-Pack $6\times 6\times6$  $k$-mesh to sample the whole Brillouin zone was critical to obtain the correct Fermi level. 
This $k$ mesh has a $k$-point ($\frac{1}{12}$, $\frac{1}{12}$, -$\frac{1}{12}$) close to the Dirac point located at ($\frac{1}{12.5}$, $\frac{1}{12.5}$, -$\frac{1}{12.5}$) in the 80-atom primitive-cell reciprocal lattice basis or at (0, 0, $\pm$0.0386 $\si{\angstrom}^{-1}$)=(0,0, $\pm\frac{1}{6.25}\frac{2\pi}{c}$) in the conventional-cell reciprocal lattice, close to the previous reports~\cite{Ali2014,Conte2017,crassee_review,Baidya2020}. 
The symmetry of \ce{Cd3As2} is $\mathrm{I4_1/acd}$ and fully exploited to minimize the computational expense. 
The relaxed lattice constants of the conventional cell are $a$=$b$=12.850\AA, $c$=25.905\AA. 
The calculated electronic structure is shown in the SI using the 80-atom primitive cell. 

In addition, we compared the experimental angle-dependent quantum oscillation frequencies with DFT calculations. With the Fourier interpolated Fermi surfaces on a $\Gamma-65\times 65 \times 83$ \textbf{k}-point gird, we used the SKEAF package~\cite{Rourke2012} to divide the Fermi surfaces into slices perpendicular to the external magnetic field and obtain the slices with local maximum or minimum areas, from which the quantum oscillation frequencies can be calculated using the Onsager relation. For evaluating thermoelectric transport coefficients, we applied the BoltzTraP2 ~\cite{Madsen2006,Madsen2018} software to calculate transport properties by solving the linearized Boltzmann transport equation for electrons under the constant relaxation time approximation. 

\section*{Data Availability}
The authors declare that the data supporting the findings of this study are available within the paper, its supplementary information files, and the Dryad Data Repository (\url{https://doi.org/10.5061/dryad.z08kprrjv}).

\begin{figure}[!htb]
\includegraphics[width=1\textwidth]{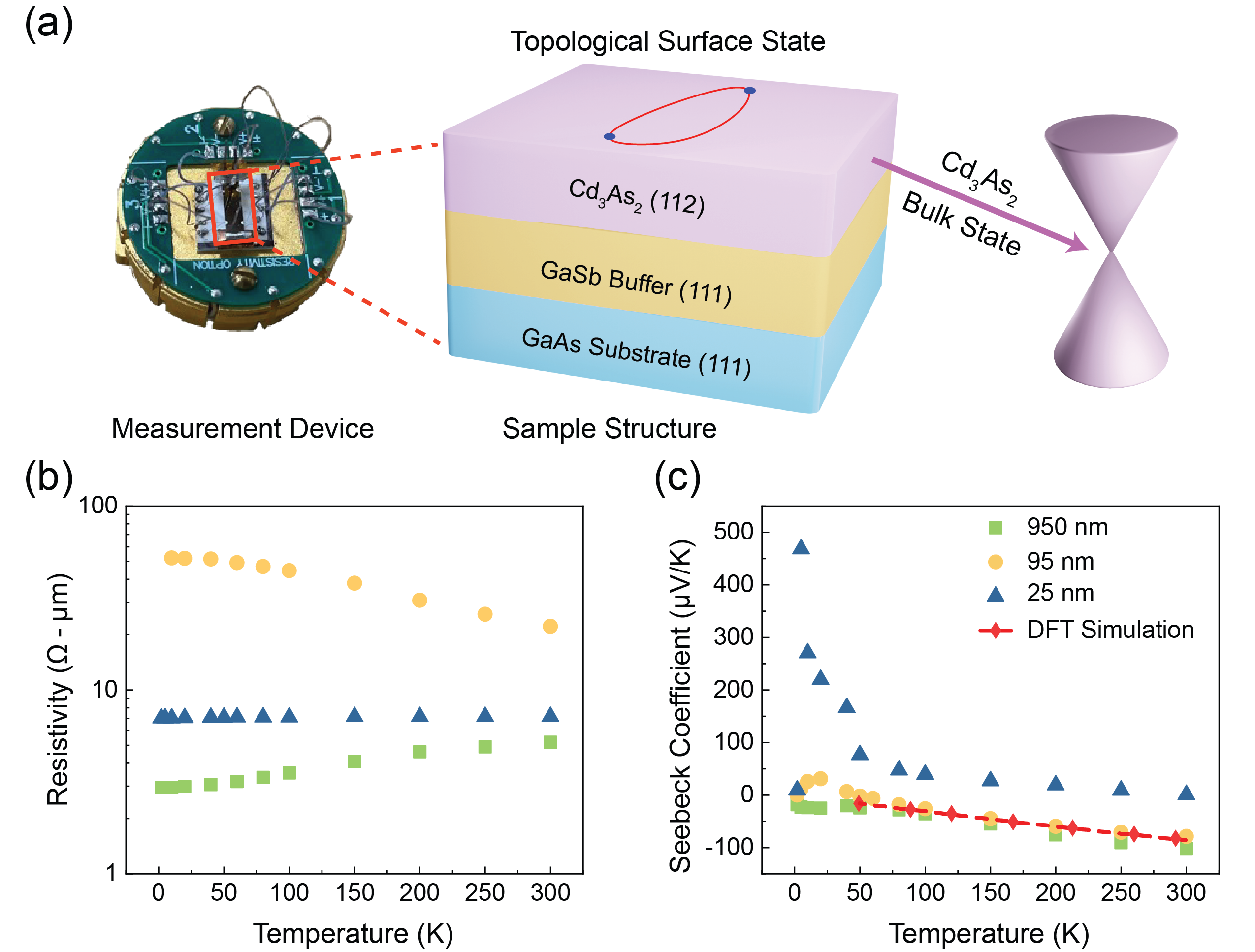}
\caption{\textbf{Sample structures and temperature dependence of resistivity and Seebeck coefficient.} (a) Schematic of the sample structure and potential conducting channels and a picture of the measurement device. (112) \ce{Cd3As2} samples with different thicknesses were epitaxially grown on a 150-nm (111) GaSb buffer layer on (111) GaAs substrates. (b) Temperature dependence of the measured resistivity of 950\,nm, 95\,nm and 25\,nm \ce{Cd3As2} samples, respectively. (c) Temperature dependence of the measured Seebeck coefficients in the three samples, as compared to DFT calculation considering only \ce{Cd3As2} bulk states.  Error bars are not shown here for clarify and are less than 10$\%$ for data shown here.} 
\label{fig:fig1}
\end{figure}

\begin{figure}[!htb]
\includegraphics[width=1\textwidth]{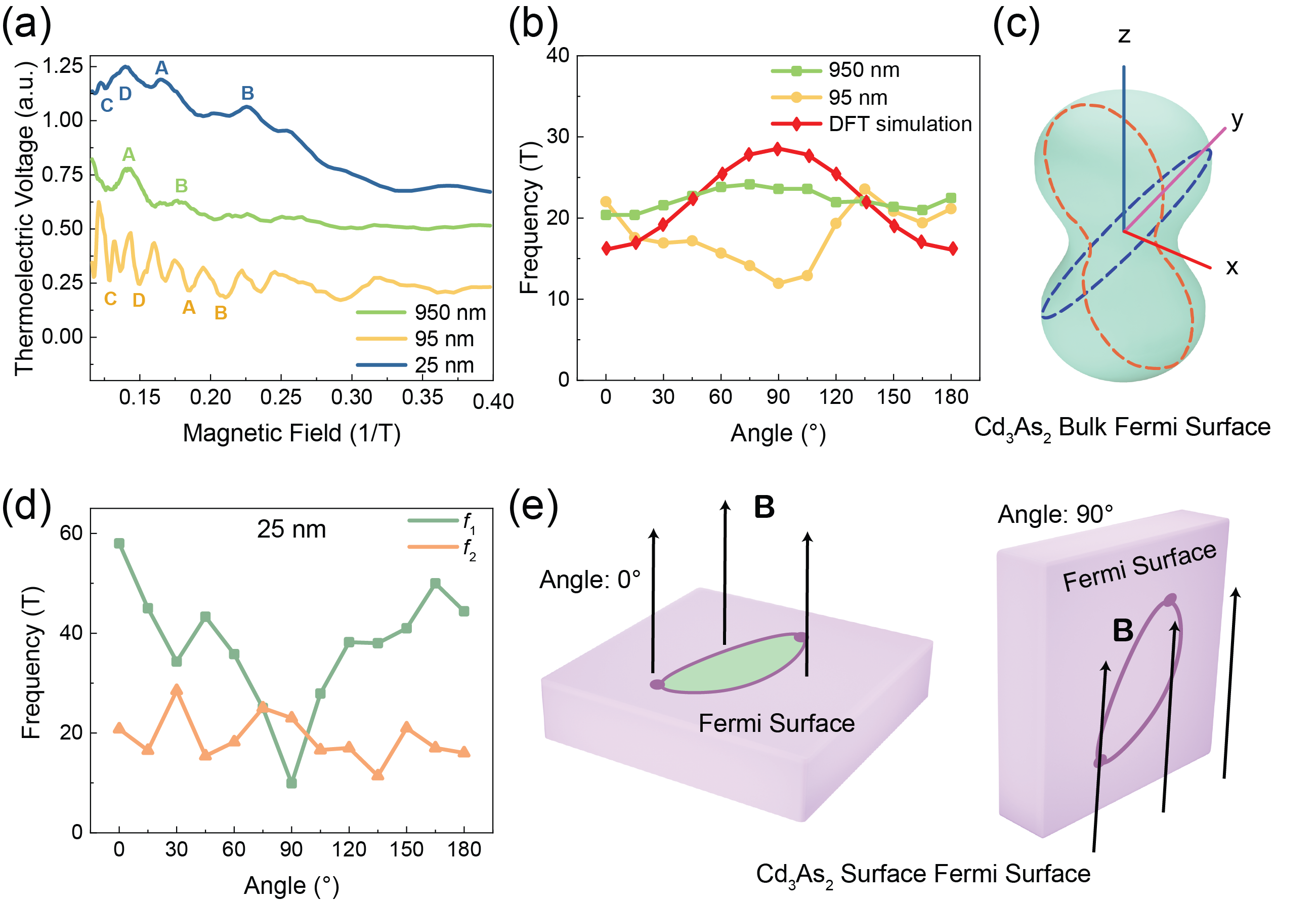}
\caption{\textbf{Angle-dependent quantum oscillations.} (a) Measured thermoelectric voltage as a function of the inverse magnetic field for the three samples.  Peaks A-B correspond to a 20\,T oscillation frequency and peaks C-D, observed only in the 95\,nm and 25\,nm samples, correspond to a 55\,T frequency. (b) Angular dependence of the quantum oscillation frequency in 950\,nm and 95\,nm samples. The oscillation frequency of these two samples is consistent with the DFT calculation considering \ce{Cd3As2} bulk states. (c) The Fermi surface of \ce{Cd3As2} bulk states responsible for the observed quantum oscillation frequency near 20\,T. (d) Angular dependence of the quantum oscillation frequency in the 25-nm sample. The oscillation frequency of the 25-nm sample exhibits two distinct frequencies. The lower one near 20\,T corresponds to the \ce{Cd3As2} bulk states. Additionally, there is a high-frequency peak near 55\,T at 0 degree, which decreases with an increasing field angle, indicating its origin in the topological surface states. (e) Illustration of the angular dependence of the oscillation frequency corresponding to topological surface states. The Fermi surface of the surface states has the maximum intersection area with the magnetic field and the highest oscillation frequency at zero degree field angle, while a 90-degree field angle parallel to the sample surface leads to a vanishing oscillation frequency. } 
\label{fig:fig2}
\end{figure}

\begin{figure}[!htb]
\includegraphics[width=1\textwidth]{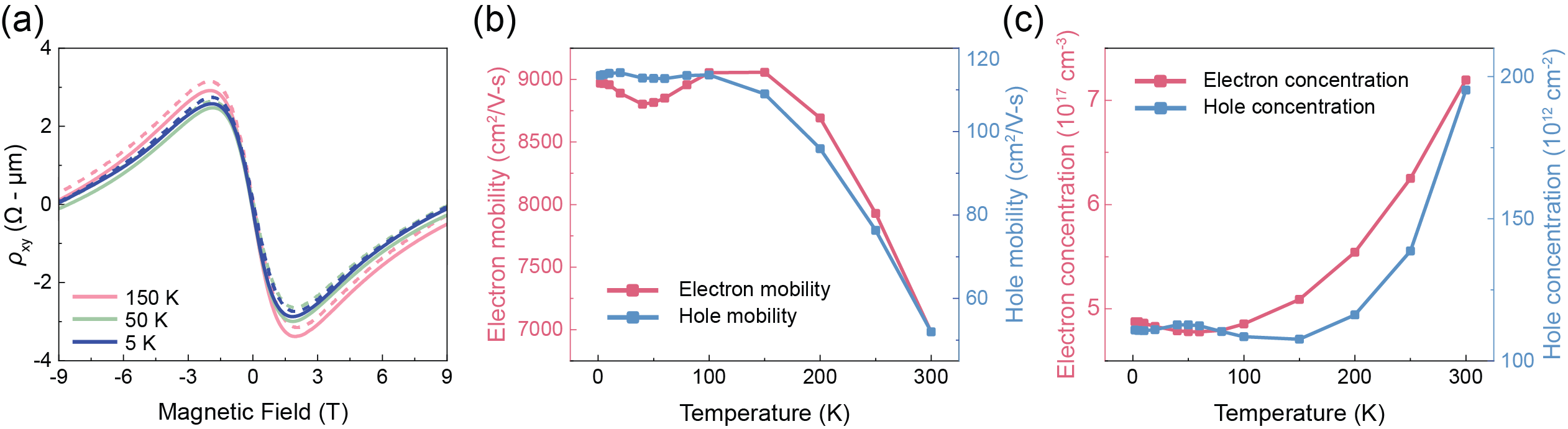}
\caption{\textbf{Magnetic-field-dependent Hall measurements of the 25-nm sample.} (a) Measured Hall resistivity (solid lines) of the 25-nm sample fitted by a two-band Drude model (dashed lines) at three representative temperatures. (b) Fitted mobilities of the electron and hole conducting channels, indicating a highly mobile n-type channel and a low-mobility p-type channel. (c) Fitted carrier concentrations of the n-type bulk states and the p-type surface states.} 
\label{fig:fig3}
\end{figure}

\begin{figure}[!htb]
\includegraphics[width=1\textwidth]{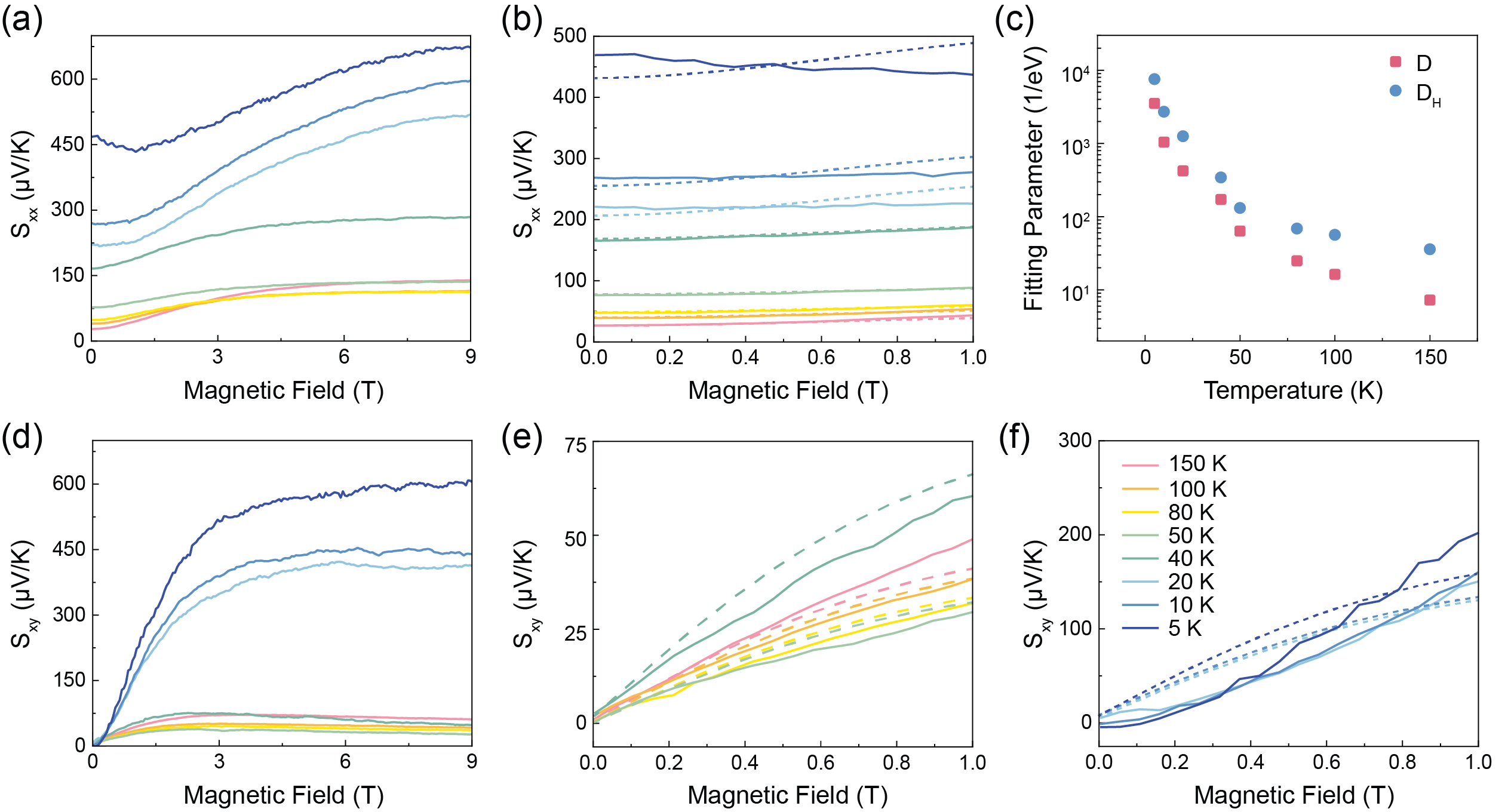}
\caption{\textbf{Magnetic-field-dependent Seebeck and Nernst measurements of the 25-nm sample.} (a) Measured Seebeck coefficient of the 25-nm sample at different temperatures with a magnetic field up to 9\,T. (b) Low-field Seebeck data (solid lines) fitted by a two-band semiclassical model (dashed lines). (c) The fitted Mott relation parameters $D$ and $D_H$ as a function of temperature. (d) Measured Nernst coefficient of the 25-nm sample at different temperatures with a magnetic field up to 9\,T. (e) Low-field Nernst data (solid lines) fitted by a two-band semiclassical model (dashed lines) above 40\,K, where reasonable agreement between experiment and model is observed. (e) Measured Nernst coefficient (solid lines) of the 25-nm sample fitted by a two-band model (dashed lines) below 20\,K, where large deviation between experiment and model potentially due to an enhanced anomalous Nernst contribution from the topological surface states is observed.} 
\label{fig:fig4}
\end{figure}

\bibliography{references.bib}

\begin{acknowledgments}
This work is based on research supported by the National Science Foundation (NSF) via the DMREF program under award number DMR-2118523. Transport measurements were done at UCSB Materials Research Laboratory (MRL) Shared Experimental Facilities, which are supported by the NSF MRSEC Program under award number DMR-2308708. The computational work used Stampede2 at Texas Advanced Computing Center (TACC) and Expanse at San Diego Supercomputer Center (SDSC) through allocation MAT200011 from the Advanced Cyberinfrastructure Coordination Ecosystem: Services \& Support (ACCESS) program, which is supported by NSF grants 2138259, 2138286, 2138307, 2137603, and 2138296. Use was also made of computational facilities purchased with funds from NSF (award number CNS-1725797) and administered by the Center for Scientific Computing (CSC) at UCSB. The CSC is supported by the California NanoSystems Institute and the UCSB MRL. 
\end{acknowledgments}

\section*{Author Contributions}

B.L., S.S. and J.P.H. conceived and supervised the project. A.C.L. prepared the samples. W.O., B.L.W., D.V. and X.L. conducted the transport measurements. Y.C. conducted the DFT simulations. H.Z. and Y.W. constructed the two-band Drude model. W.O., Y.C. and B.L. drafted the manuscript. All authors have commented and edited the manuscript. 

\section*{Competing Interests}

The authors declare no competing interests.

\end{document}



\title{Supplementary Information: Extraordinary Thermoelectric Properties of Topological Surface States in Quantum-Confined \ce{Cd3As2} Thin Films} 

\author{Wenkai Ouyang}
\affiliation{Department of Mechanical Engineering, University of California, Santa Barbara, CA 93106, USA}

\author{Alexander C. Lygo}
\affiliation{Materials Department, University of California, Santa Barbara, CA 93106, USA}

\author{Yubi Chen}
\affiliation{Department of Mechanical Engineering, University of California, Santa Barbara, CA 93106, USA}
\affiliation{Department of Physics, University of California, Santa Barbara, CA 93106, USA}

\author{Huiyuan Zheng}
\affiliation{Department of Physics, the University of Hong Kong, Hong Kong, 999077, China}

\author{Dung Vu}
\affiliation{Department of Mechanical and Aerospace Engineering, the Ohio State University, Columbus, OH 43210, USA}

\author{Brandi L. Wooten}
\affiliation{Department of Mechanical and Aerospace Engineering, the Ohio State University, Columbus, OH 43210, USA}

\author{Xichen Liang}
\affiliation{Department of Chemical Engineering, University of California, Santa Barbara, CA 93106, USA}

\author{Wang Yao}
\affiliation{Department of Physics, the University of Hong Kong, Hong Kong, 999077, China}

\author{Joseph P. Heremans}
\email{heremans.1@osu.edu}
\affiliation{Department of Mechanical and Aerospace Engineering, the Ohio State University, Columbus, OH 43210, USA}

\author{Susanne Stemmer}
\email{stemmer@ucsb.edu}
\affiliation{Materials Department, University of California, Santa Barbara, CA 93106, USA}

\author{Bolin Liao}
\email{bliao@ucsb.edu} \affiliation{Department of Mechanical Engineering, University of California, Santa Barbara, CA 93106, USA}

\date{\today}
                            
\maketitle


\section*{Supplementary Note 1: Bare Gallium Antimonide Layer Characterization}
\begin{figure}[!htb]
\includegraphics[width=0.75\textwidth]{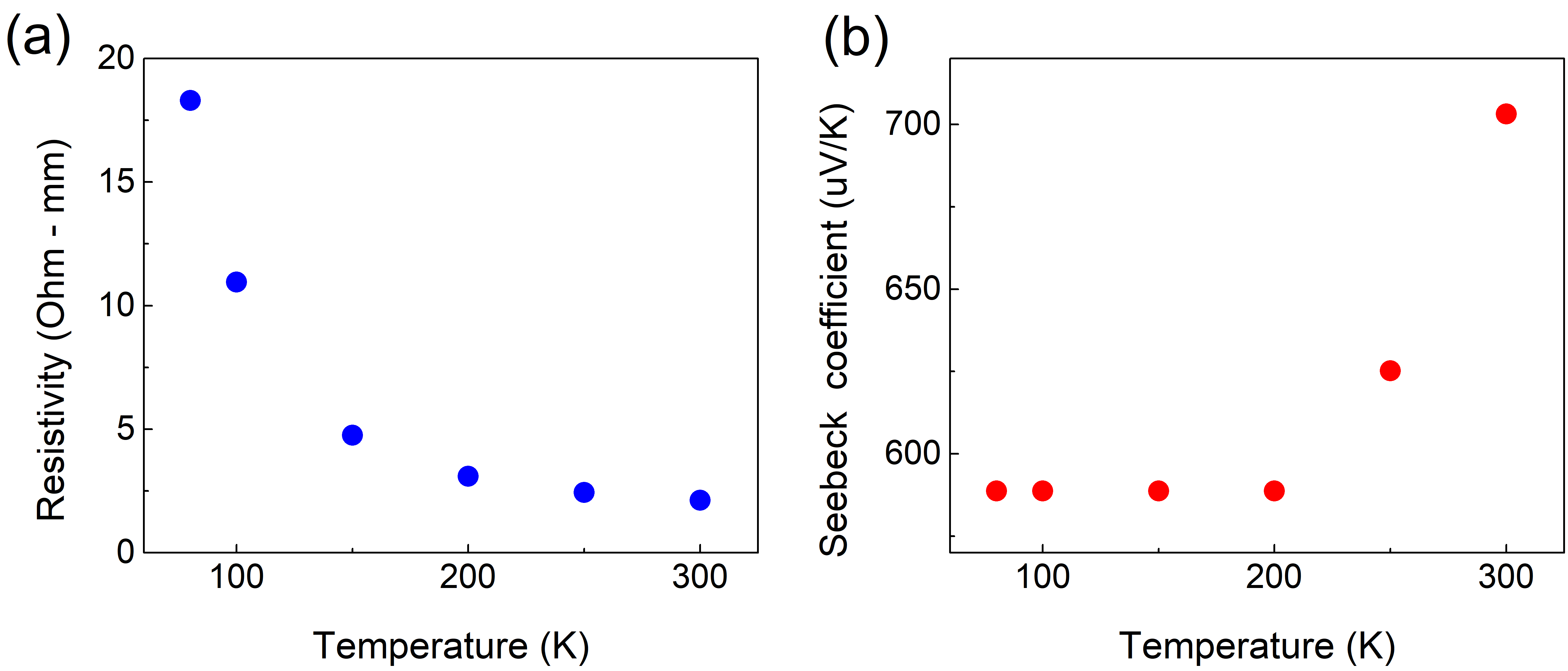}
\caption{\textbf{Temperature dependence of resistivity and Seebeck coefficient of a bare GaSb buffer layer.} (a) Temperature dependence of the electrical resistivity of a 150-nm GaSb buffer layer grown on a GaAs substrate. The buffer layer becomes insulating below 50\,K. (c) Temperature dependence of the Seebeck coefficient of a 150-nm GaSb buffer layer grown on a GaAs substrate.} 
\label{fig:GaSb}
\end{figure}

The electrical transport properties and thermoelectric properties of a bare 150-nm GaSb layer grown on a GaAs substrate using MBE were measured using our steady-state transport measurement system. The results are shown in Fig.~\ref{fig:GaSb}. The resistivity of GaSb is found to be on the order of $10^{-3}\ \Omega\text{-}\text{m}$, while the resistivity of \ce{Cd3As2} is on the order of $10^{-6}\ \Omega\text{-}\text{m}$ at room temperature. The GaSb layer becomes insulating below 50\,K. The Seebeck measurement indicates the prevalence of p-type carriers in GaSb, with a Seebeck coefficient of approximately 700\,$\mu \mathrm{V/K}$ at room temperature, which decreases with a decreasing temperature. 

\clearpage

\section*{Supplementary Note 2: Additional magnetotransport data and Analytical Models}
\subsection{Field-dependent Electrical Resistivity and Hall Measurement}
\begin{figure}[!htb]
\includegraphics[width=1.0\textwidth]{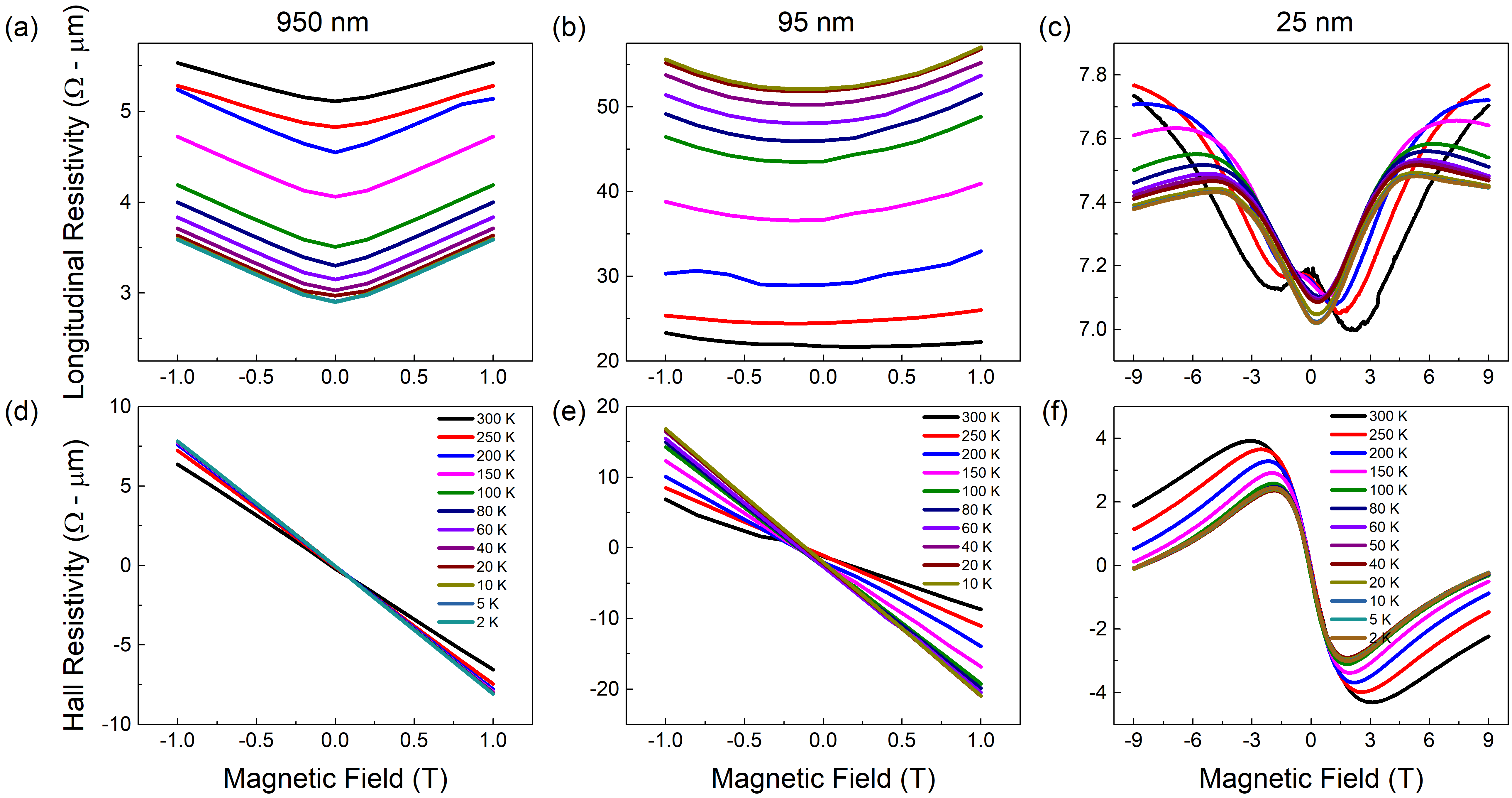}
\caption{\textbf{Field dependence of longitudinal resistivity and Hall resistivity measured at different temperatures.} (a-c) The longitudinal resistivity of 950 nm, 95 nm and 25 nm \ce{Cd3As2}, respectively. (d-f) The Hall resistivity of the three samples.} 
\label{fig:figS2}
\end{figure}

We conducted magneto-electrical transport measurements to determine the longitudinal and Hall resistivity of the three \ce{Cd3As2} samples. The obtained results are shown in Fig.~\ref{fig:figS2}. As for 950 nm and 95 nm \ce{Cd3As2} samples, the electrical measurements were carried out under a magnetic field strength up to 1\,T only to exact the low-field transport properties, while the 25 nm \ce{Cd3As2} sample was subjected to a magnetic field strength up to 9\,T. The determination of carrier type was based on the slope of  the low-field Hall resistivity. The low-field Hall resistivity reveals that both the 950 nm and 95 nm samples have dominant n-type carriers, which are consistent with the results obtained from our Seebeck measurements. In the case of the 25 nm \ce{Cd3As2} sample, the change of the Hall resistivity slope at higher magnetic field indicates the existence of both n-type and p-type carriers. The negative slope at small magnetic field signals that the n-type carriers have a higher mobility, while the positive slope at higher magnetic field suggests the p-type carriers have a higher density. Furthermore, the dissimilar shapes observed in the longitudinal and transverse resistivity emphasize distinct transport characteristics, requiring the adoption of different models for their respective fitting analyses.

\clearpage

\subsection{Analytical Magneto-electrical transport Models}
\begin{figure}[!htb]
\includegraphics[width=1.0\textwidth]{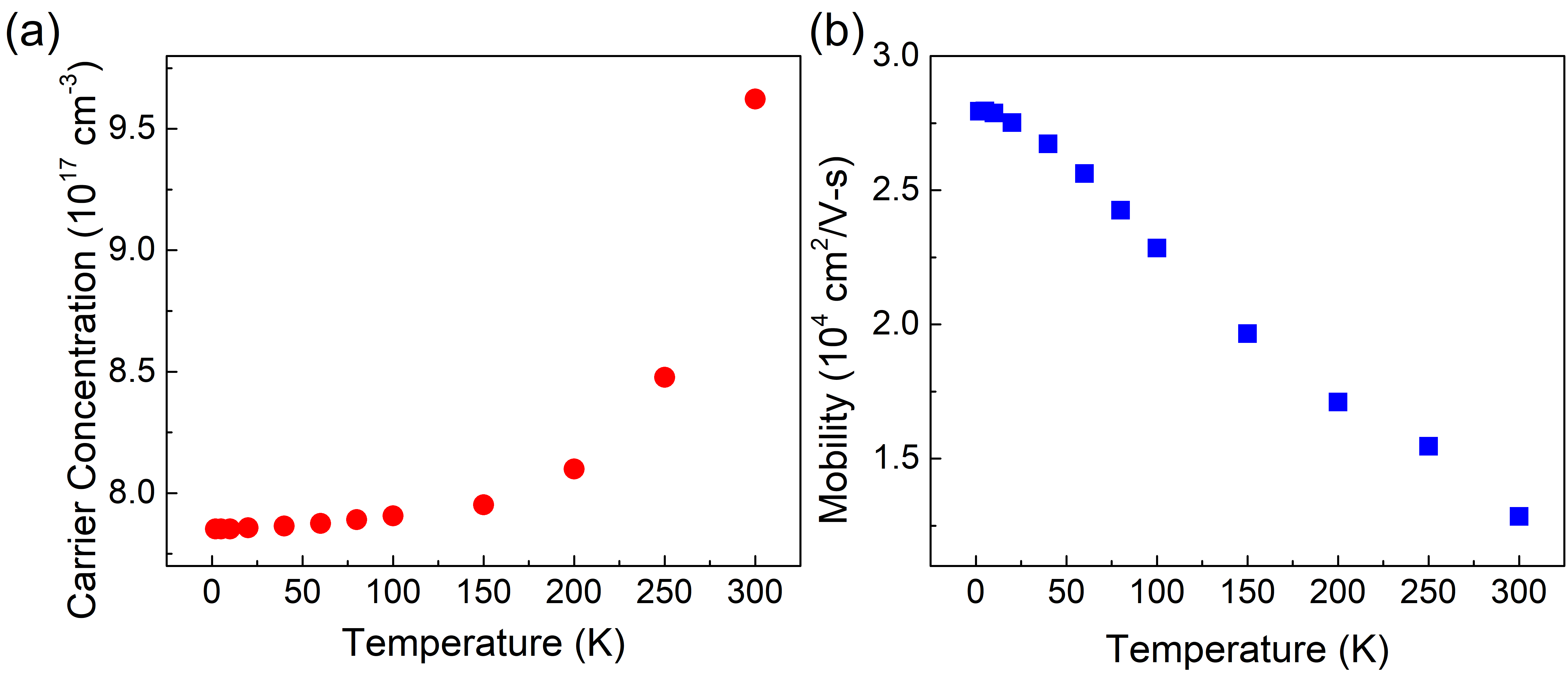}
\caption{\textbf{Carrier concentration and mobility in the 950 nm sample extracted from the Hall resistivity measurement.} (a) The fitted carrier concentration in the 950 nm sample at different temperatures. (b) The fitted carrier mobility in the 950 nm sample at different temperatures.} 
\label{fig:950nm}
\end{figure}

Based on the quantum oscillation analysis, it is established that the dominant carrier in the 950 nm \ce{Cd3As2} sample corresponds to the n-type bulk states, whereas in the 25 nm \ce{Cd3As2} sample, two groups of carriers associated with bulk states and topological surface states contribute to the transport properties. This interpretation is also supported by the Hall resistivity measurements shown in Fig.~\ref{fig:figS2}. Specifically, in the 950 nm sample, the Hall resistivity exhibits a linear response with an increasing magnetic field, while in the 25 nm sample, it follows a nonlinear trend, indicating the contributions from two carriers. Consequently, we construct a one-band and a two-band Drude model to describe transport in the 950 nm and the 25 nm samples, respectively. For the 950 nm \ce{Cd3As2} sample, its resistivity tensor 
$\rho =  
    \begin{pmatrix}
        \rho_{xx} & -\rho_{yx} \\
        \rho_{yx} & \rho_{yy}
    \end{pmatrix}$
is described by $\rho_{xx} = \rho_{yy} = 1/(n_e e \mu_e)$ and $\rho_{yx} = B/(n_e e)$
where $e$ is the elementary charge, $n_e$ is the carrier density of the bulk states, and $\mu_e$ is the mobility of the bulk states~\cite{ashcroft1976solid}. The fitted carrier concentration and mobility in the 950 nm sample at different temperatures is shown in Fig.~\ref{fig:950nm}.

For the 25 nm \ce{Cd3As2} sample, the resistivity tensor is described by contributions from the surface carriers (labeled by subscript $s$) and the bulk carrier (labeled by subscript $b$), whose longitudinal resistivity and Hall coefficient are denoted by $\rho_{s/b}$ and $R_{s/b}$, respectively:
\begin{eqnarray*}
    \rho_{xx} &=& \frac{\rho_s \rho_b (\rho_s + \rho_b) + (\rho_s R_b^2 + \rho_b R_s^2)B^2}{(\rho_s + \rho_b)^2 + (R_s + R_b)^2B^2}, \\
    \rho_{yx} &=& \frac{(R_s \rho_b^2 + R_b \rho_s^2)B + R_s R_b (R_s + R_b)B^3}{(\rho_s + \rho_b)^2 + (R_s + R_b)^2B^2}.
\end{eqnarray*}
The parameters obtained from the transverse resistivity $\rho_{yx}$ for the two carriers often fail to accurately replicate the longitudinal resistivity $\rho_{xx}$ due to the absence of consideration for the magnetic-field dependence of the relaxation time in the model~\cite{ando2013topological}. However, we can impose constraints on our fitting by verifying the consistency of $\rho_{xx}$ at low magnetic fields ($<$ 1\,T). With the above constraints, we perform the fitting of the Hall conductivity of the 25 nm sample, as shown in Fig.~3 of the main text. The transport properties exhibit a discernible influence from surface states, which is evidenced by the presence of p-type surface carriers. 


\clearpage




\subsection{Magneto-thermoelectric Measurements of 950 nm and 95 nm samples}

\begin{figure}[!htb]
\includegraphics[width=1\textwidth]{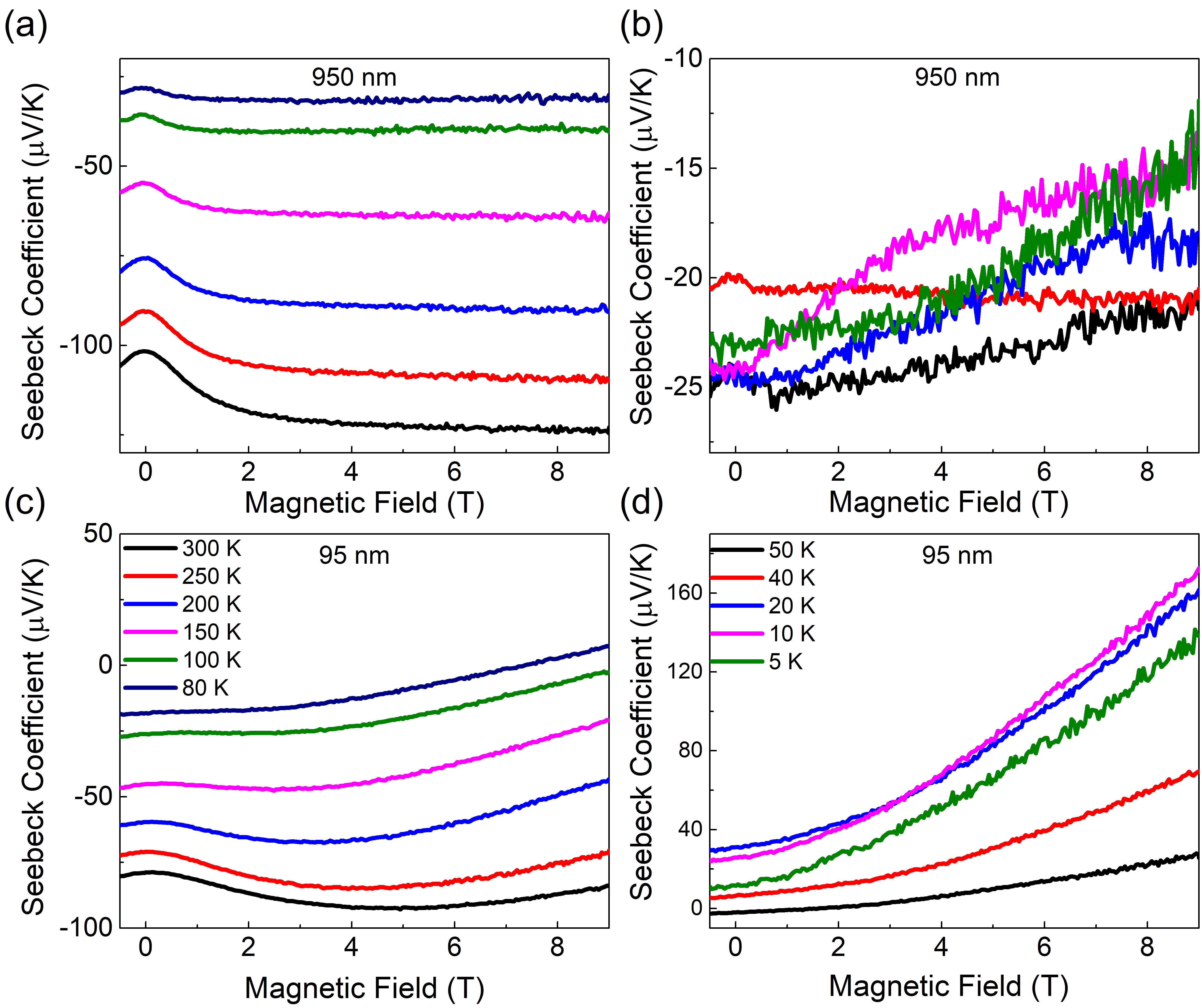}
\caption{\textbf{Magnetic field dependence of the Seebeck coefficient in 950 nm and 95 nm sample at different temperatures.} (a-b) The Seebeck coefficient measured in the 950 nm sample. (c-d) The Seebeck coefficient measured in the 95 nm sample. } 
\label{fig:figS3}
\end{figure}

\begin{figure}[!htb]
\includegraphics[width=1\textwidth]{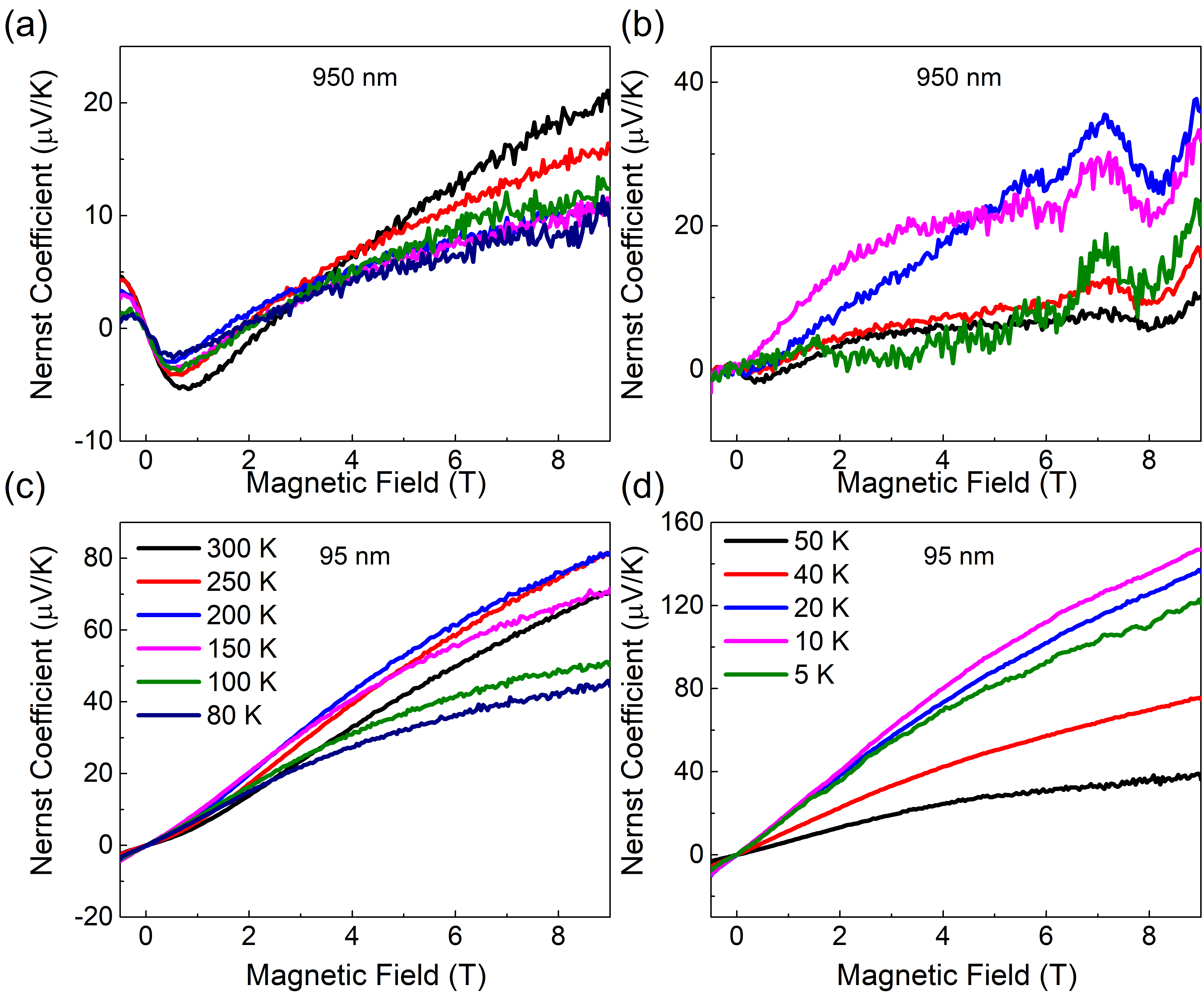}
\caption{\textbf{Magnetic field dependence of the Nernst coefficient in 950 nm and 95 nm sample at different temperatures.} (a-b) The Nernst coefficient measured in the 950 nm sample. (c-d) The Nernst coefficient measured in the 95 nm sample. } 
\label{fig:figS4}
\end{figure}
\clearpage

\subsection{Fitting magneto-thermoelectric data of the 25 nm sample using experimental conductivity values}
\begin{figure}[!htb]
\includegraphics[width=1.05\textwidth]{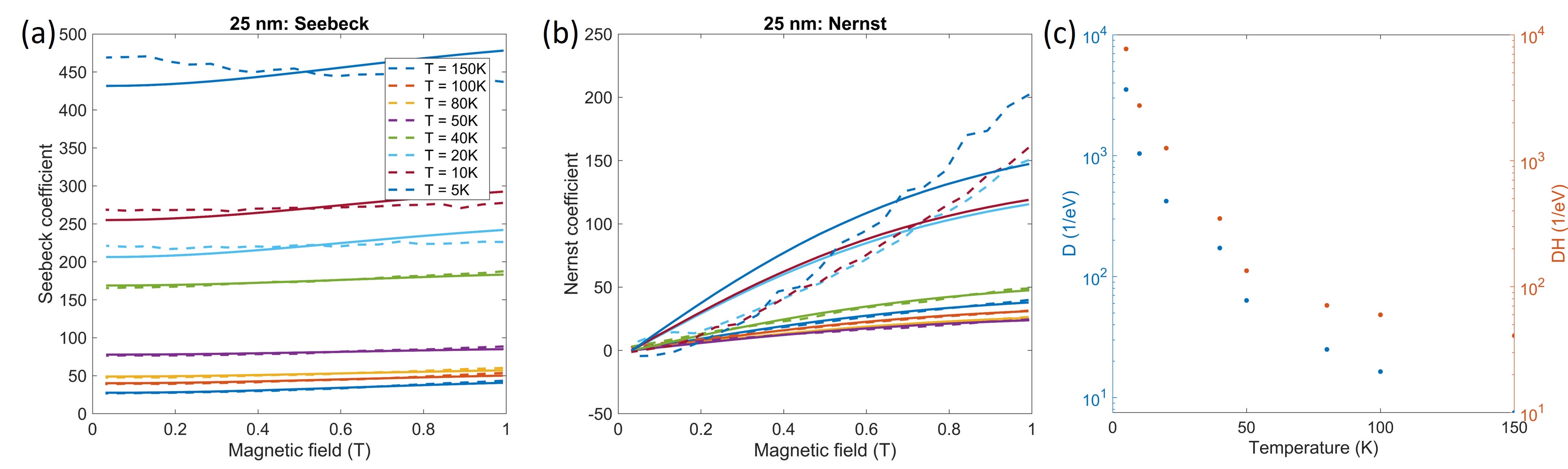}
\caption{\textbf{Magneto-thermoelectric data of the 25 nm sample fitted using Eqn.~(1) and (2) in the main text and experimental conductivity values.} (a) Experimental and fitted magneto-Seebeck coefficient of the 25 nm sample. (b) Experimental and fitted magneto-Nernst coefficient of the 25 nm sample. (c) Fitting parameters $D$ and $D_H$ for magnetothermoelectric transport in the 25 nm sample. The fitting results and the fitting parameter values are similar to those shown in Fig.~4 of the main text.} 
\label{fig:figS5}
\end{figure}
\clearpage

\section*{Supplementary Note 3: Additional Computational Methods and Data}

\begin{figure}[!htb]
\includegraphics[width=1.0\textwidth]{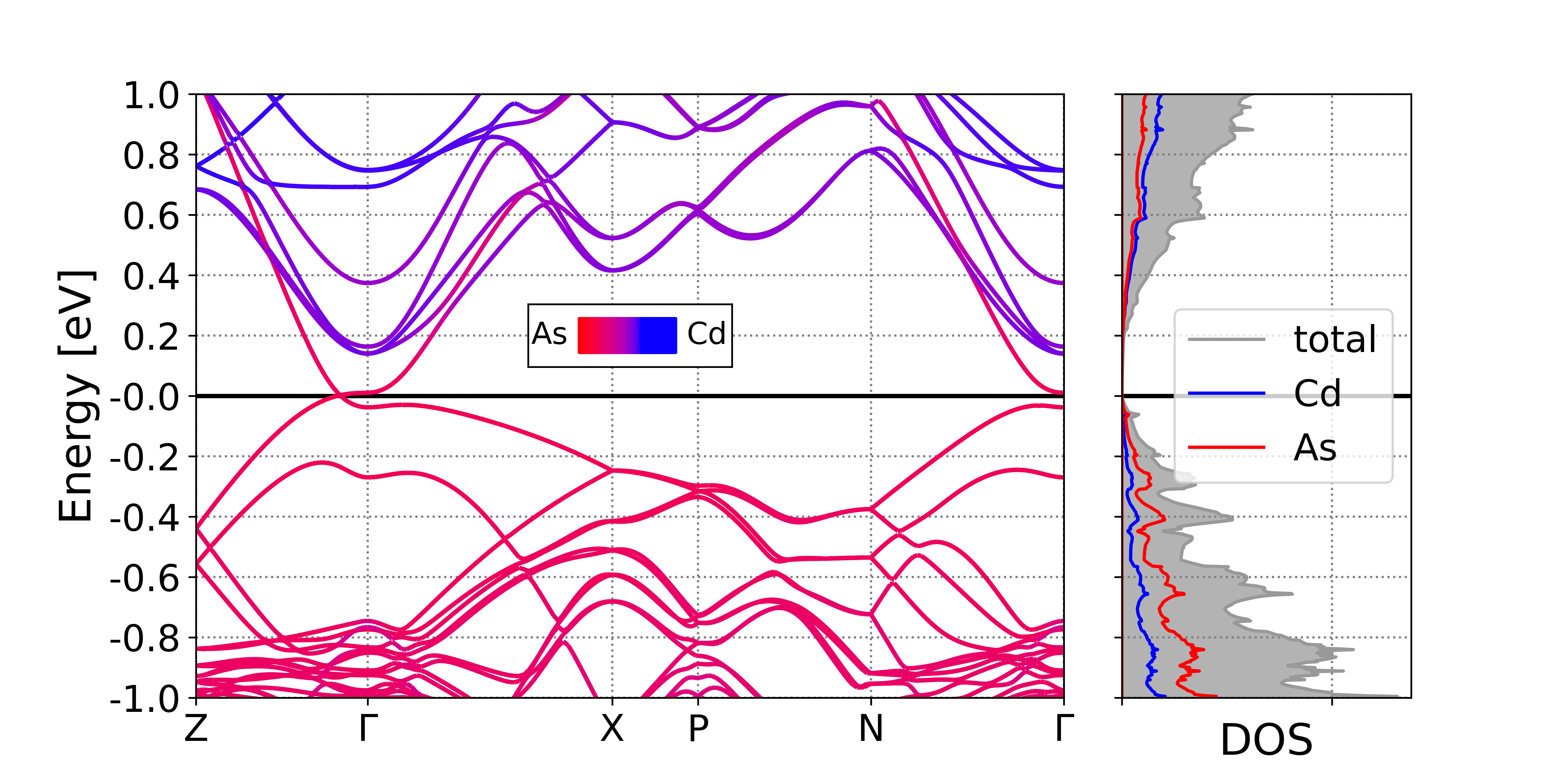}
\caption{
\textbf{\ce{Cd3As2} electron band structure and density of states (DOS)}. 
The variations in color indicate the atomic character of the electron states, as represented on the associated color bar.
The energy reference is the Dirac crossing point.
}
\label{fig:figS6}
\end{figure}

\begin{figure}[!htb]
\includegraphics[width=1.0\textwidth]{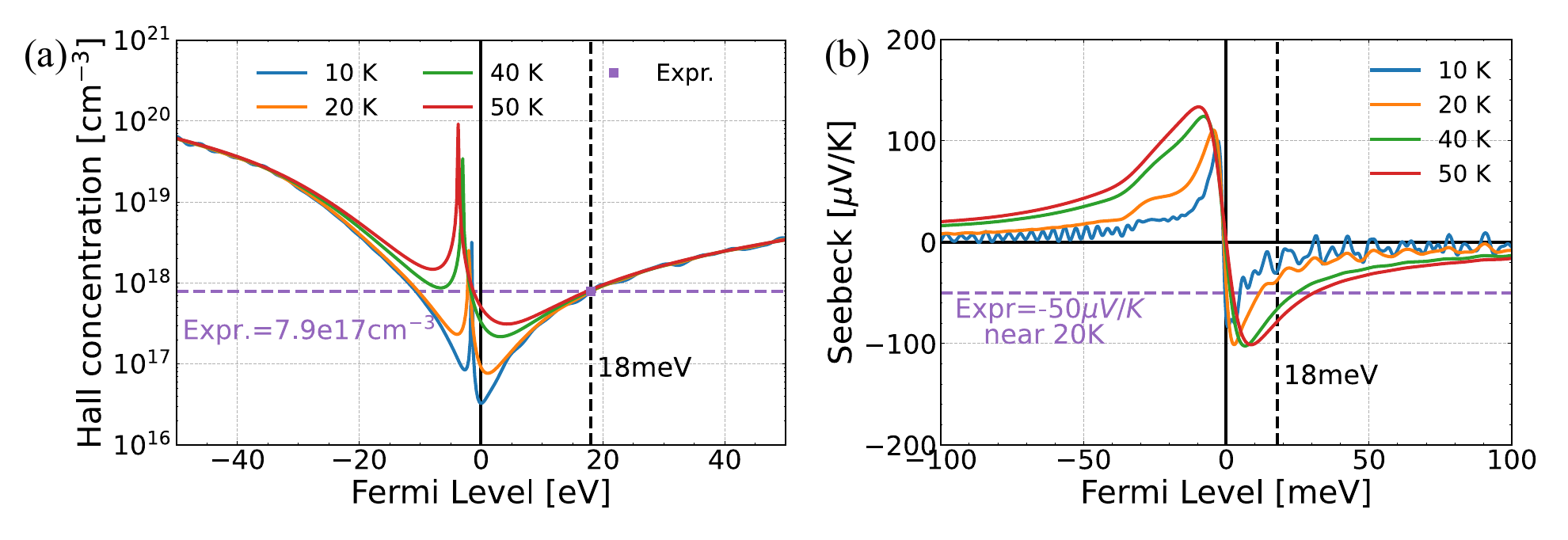}
\caption{
The calculated (a) Hall concentration and (b) Seebeck coefficients in bulk \ce{Cd3As2} as a function of Fermi level under low temperatures (10~K, 20~K, 40~K, and 50~K).
The purple dashed line indicates the experimental values at low T. The Hall concentration and Seebeck coefficient can be nearly matched with the same Fermi level at 18~meV.
}
\label{fig:figS7}
\end{figure}

\begin{figure}[!htb]
\includegraphics[width=0.9\textwidth]{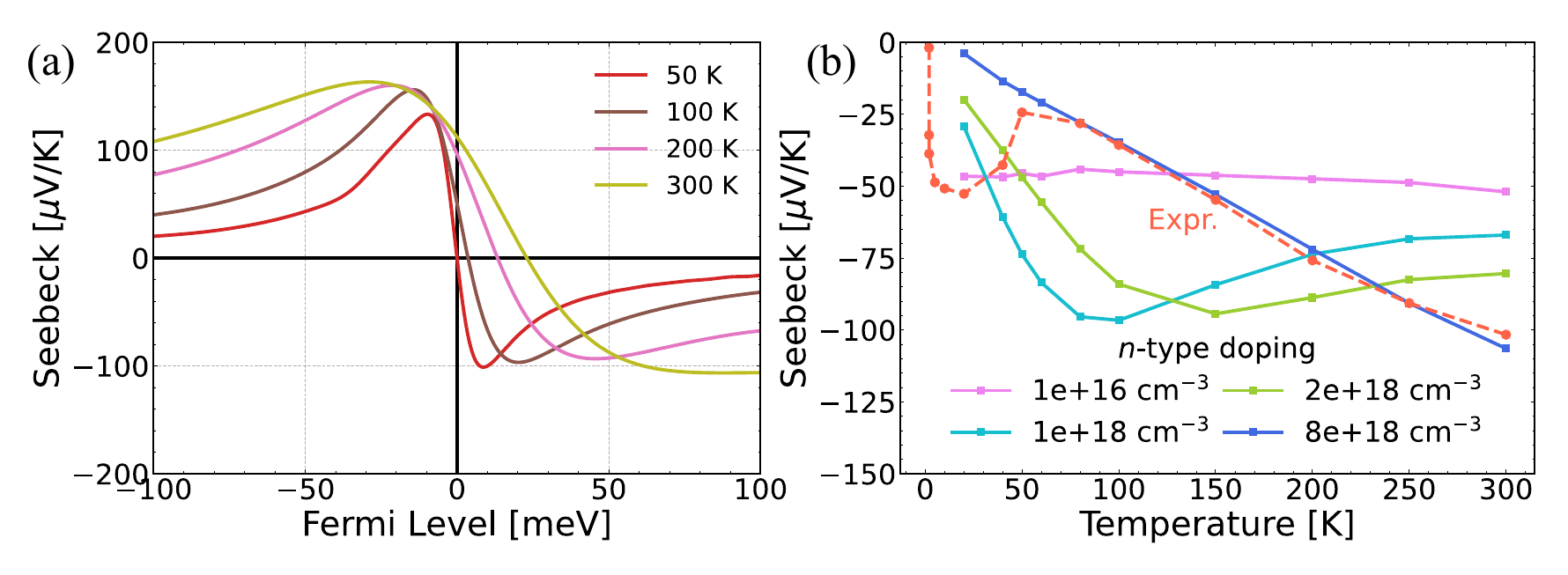}
\caption{
(a) The calculated Seebeck coefficient across Fermi level at higher temperatures (50~K, 100~K, 200~K, 300~K).
(b) The calculated Seebeck coefficient of bulk \ce{Cd3As2} with different n-type carrier concentrations and temperatures.
}
\label{fig:figS8}
\end{figure}

DFT calculations can provide the basis to determine electron transport and other physical observables like the Seebeck coefficient. The parameters for the ground state DFT calculation are provided in the Methods section in the Main Text. The calculated electronic band structure of \ce{Cd3As2} is shown in Fig.~\ref{fig:figS6}.
To evaluate the transport properties, we apply the BoltzTraP2~\cite{Madsen2006,Madsen2018} software, which allows for the calculation of transport properties by solving the linearized Boltzmann equation under the constant relaxation time approximation. 
Of the various physical observables considered, the Hall carrier concentration in Fig.~\ref{fig:figS7}(a) and the Seebeck coefficient (shown in Fig.~1c in the Main Text, Fig.~\ref{fig:figS7}(b), and Fig.~\ref{fig:figS8}) are independent from the choice of the relaxation time, and they are estimated with respect to the Fermi level position at different temperatures. 
The \ce{Cd3As2} energy surfaces are obtained from a non-self-consistent calculation on a $\Gamma$-$12\times12\times12$ \textbf{k}-point grid and subsequently Fourier interpolated~\cite{Madsen2018} onto a dense $\Gamma$-$65\times65\times83$ \textbf{k}-point grid. 
The dense \textbf{k}-point grid is also necessary to ensure the convergence of the Seebeck and Hall concentration calculation.
The experimental Hall measurement of the 950 nm sample resulted in a Hall concentration of approximately $7.9\times10^{17}$ $\mathrm{cm}^{-3}$ at low temperatures in Fig.~\ref{fig:950nm}(a), which corresponds to a Fermi level near 18~meV above the charge neutrality point from our DFT simulation in Fig.~\ref{fig:figS7}(a). 
In Fig.~\ref{fig:figS7}(b), the experimental Seebeck coefficient of the 950~nm sample is around $-50\mu V/K$, close to the Seebeck coefficients at the 18~meV Fermi level pinned by the Hall measurement.
Our findings demonstrate that the predominant thermoelectric transport in the 950 nm sample stems from the bulk states, as confirmed by the consistency of our DFT calculations with the experimental results.

For the thermoelectric transport simulation, the Mott formula~\cite{ashcroft1976solid} is used to numerically evaluate the Seebeck coefficient in Fig.~\ref{fig:figS7}(b) and Fig.~\ref{fig:figS8}. 
In Fig.~\ref{fig:figS8}(b), the low temperature experimental Seebeck coefficient almost shares the same Fermi level as Hall concentration in Fig.~\ref{fig:figS7}.
As for the high temperature Seebeck coefficient (above 50~K), the trend is well captured by $n$-type doping concentration $8\times10^{18}~\mathrm{cm^{-3}}$.
We estimated the experimental Seebeck coefficient of the 950 nm sample with different $n$-type carrier concentrations. 
The numerical determination of the Fermi level at various temperatures and carrier concentrations is achieved through the utilization of the charge neutrality equation, which then enables the corresponding Seebeck coefficient to be extracted with respect to the obtained Fermi level. 
The resulting Seebeck data in Fig.~\ref{fig:figS8} with $n$-type $8\times10^{18}~\mathrm{cm^{-3}}$ carrier concentration at higher temperatures is found to exhibit the best agreement with the 950 nm experimental results, although the carrier concentration in the 950~nm sample extracted from the Hall resistivity measurement is around $7.9\times 10^{17}$ $\mathrm{cm^{-3}}$.
This discrepancy can be attributed to the limited accuracy of the DFT electronic structure given the small energy scale associated with the semimetallic states. 
Besides, for a complex material like \ce{Cd3As2} with a 80-atom unit cell, the computational cost limits the exploration for finite-temperature simulations, and more accurate exchange correlation functionals.
As a result, the low temperature Seebeck coefficients are well captured by the almost undoped $1\times10^{16}~\mathrm{cm^{-3}}$, and the high temperature Seebeck coefficients can be well approximated by the $8\times10^{18}~\mathrm{cm^{-3}}$ $n$-type doping concentration.

\clearpage

\textbf{Supplementary References}
\bibliography{references.bib}